\documentclass{aastex}
\usepackage{emulateapj5}
\submitted{accepted for publication in the Astrophysical Journal (Letters)}

\shorttitle{Optical Spectropolarimetry of SN~2002ap}
\shortauthors{Kawabata et al.}

\begin{document}

\title{Optical Spectropolarimetry of SN~2002ap: A
High Velocity Asymmetric Explosion\footnotemark[1]}

\footnotetext[1]{Based on data obtained at the Subaru Telescope,
which is operated by the National Astronomical Observatory of
Japan (NAOJ)}

\author{K.~S.~Kawabata\altaffilmark{2,3},
       D.~J.~Jeffery\altaffilmark{4},
       M.~Iye\altaffilmark{2,5},
       Y.~Ohyama\altaffilmark{6},
       G.~Kosugi\altaffilmark{6},
       N.~Kashikawa\altaffilmark{2},
       N.~Ebizuka\altaffilmark{7},
       T.~Sasaki\altaffilmark{6},
       K.~Sekiguchi\altaffilmark{6},
       K.~Nomoto\altaffilmark{8,9},
       P.~Mazzali\altaffilmark{9,8,10},
       J.~Deng\altaffilmark{9,8},
       K.~Maeda\altaffilmark{8},
       K.~Aoki\altaffilmark{6},
       Y.~Saito\altaffilmark{2},
       T.~Takata\altaffilmark{6},
       M.~Yoshida\altaffilmark{11},
       R.~Asai\altaffilmark{8},
       M.~Inata\altaffilmark{11},
       K.~Okita\altaffilmark{11},
       K.~Ota\altaffilmark{8,2},
       T.~Ozawa\altaffilmark{12},
       Y.~Shimizu\altaffilmark{11},
       H.~Taguchi\altaffilmark{13},
       Y.~Yadoumaru\altaffilmark{12},
       T.~Misawa\altaffilmark{8,2},
       F.~Nakata\altaffilmark{8,2},
       T.~Yamada\altaffilmark{2},
       I.~Tanaka\altaffilmark{2},
       and
       T.~Kodama\altaffilmark{14}
}

       \altaffiltext{2}{Opt. \& IR Astron. Div., NAOJ, Mitaka,
        Tokyo 181-8588, Japan}
       \altaffiltext{3}{E-mail: koji.kawabata@nao.ac.jp}
       \altaffiltext{4}{Dep. of Phys., New Mexico Institute of
        Mining and Technology, Socorro, NM 87801, USA}
       \altaffiltext{5}{Dep. of Astron., Graduate Univ.
        for Advanced Studies, Mitaka, Tokyo 181-8588, Japan}
       \altaffiltext{6}{Subaru Telescope, NAOJ, 650 North A'ohoku Place,
        Hilo, HI 96720, USA}
       \altaffiltext{7}{RIKEN, Wako, Saitama 351-0198, Japan}
       \altaffiltext{8}{Dep. of Astron., Univ. of Tokyo,
       Bunkyo-ku, Tokyo 113-0033, Japan}
       \altaffiltext{9}{Research Center for the Early Universe,
        Univ. of Tokyo, Bunkyo-ku, Tokyo 113-0033, Japan}
       \altaffiltext{10}{Osservatorio Astronomico, Via Tiepolo,
        11, 34131 Trieste, Italy}
       \altaffiltext{11}{Okayama Astrophys. Obs., NAOJ,
        Asakuchi-gun, Okayama 719-0232, Japan}
       \altaffiltext{12}{Misato Obs., Amakusa-gun, Wakayama
       640-1366, Japan}
       \altaffiltext{13}{Dep. of Astron. \& Earth Sci.,
        Tokyo Gakugei University, Koganei, Tokyo 184-8501, Japan}
       \altaffiltext{14}{Theory Div., NAOJ, Mitaka, Tokyo 181-8588, Japan}

\begin{abstract}
We present spectropolarimetry of the Type Ic supernova SN 2002ap
and give a preliminary analysis:
the data were taken at two epochs, close to and one month later than
the visual maximum (2002 February 8).
In addition we present June 9 spectropolarimetry without analysis.
The data show the development of linear polarization.
Distinct polarization profiles were seen only in the
\ion{O}{1}~$\lambda 7773$ multiplet/\ion{Ca}{2}~IR triplet
absorption trough at maximum light and in the 
\ion{O}{1}~$\lambda 7773$ multiplet and
\ion{Ca}{2}~IR triplet absorption troughs
a month later, with the latter showing
a peak polarization as high as $\sim 2$ \%.
The intrinsic polarization shows three clear position angles:
$80\arcdeg$ for the February continuum, $120\arcdeg$ for the
February line feature, and $150\arcdeg$ for the March data.
We conclude that there are multiple asymmetric components
in the ejecta.
We suggest that the supernova has a bulk asymmetry with
an axial ratio projected on the sky that is different
from 1 by of order $10$ \%.
Furthermore, we suggest very speculatively that a high
velocity ejecta component moving faster than $\sim 0.115c$
(e.g., a jet) contributes to polarization
in the February epoch.
\end{abstract}

\keywords{polarization --- supernovae: individual (SN 2002ap)}

\section{INTRODUCTION}

SN~2002ap was discovered in the nearby spiral galaxy M74 ($=$ NGC 628)
on 29 January 2002
\citep{nak02} and reached its maximum of $V\sim 12.4$ mag on
February 8 \citep{gal02}.
It has been classified as a Type~Ic supernova (SN~Ic) and
suggested to be a hypernova (but at the low-energy end of the
sequence of hypernovae;
% from the fact that its early spectra
%had very broad absorption lines
\citealt{maz02} and references therein).
%Because of its apparent optical brightness, SN~2002ap provided us
%with a rare opportunity to carry out multi-epoch,
%high-quality spectropolarimetry of a peculiar supernova.
%Unfortunately, the supernova was lost behind the Sun from
%mid-March to early June, limiting our ability to observe
%it in the brightest phase.

%\section{SNe~IC, HYPERNOVAE, SUPERNOVA POLARIZATION}

A SN Ic is thought to be the result of the core collapse of a
massive star that has either lost its hydrogen and helium envelopes
prior to the explosion or has an invisible helium envelope due to
low excitation.
The details of the explosion mechanism are still
under discussion (\citealt{nom95,bra01} and references
therein).
%
%It has recently been recognized that there is a subgroup
%of SNe Ic whose members, tentatively called hypernovae,
%exhibit very broad absorption lines in their early spectra.
%Some hypernovae have had their spectra successfully modeled as
%the hyperenergetic explosion of a massive C$+$O star,
%with an explosive kinetic energy exceeding $\sim 5$--$10$ times as
%much as that of normal core-collapse SNe
%(\citealt{iwa98,nom01} and references therein).
SN 1998bw, the most luminous and energetic
Type Ic
`hypernova' to date,
has been particularly well studied, and its
probable connection with the $\gamma$-ray burst GRB 980425 has been
pointed out (e.g., \citealt{gal98,iwa98,nom01}).
An aspherical hyperenergetic explosion has been suggested
to explain the
slowly-declining light curve of SN 1998bw and the narrowness
of the [\ion{O}{1}] $\lambda$6300, 6363 emission line in the
nebular phase \citep{maz01,nak01,mae02}.
%Alternatively, \citet{hoe99} suggested that the observed behavior
%could be explained by a moderate explosion
%($2\times 10^{51}$ ergs) if the ejecta had a prolate asphericity
%with an axial ratio of about 2 and were viewed close to the symmetry
%axis.

Intrinsic polarization is zero for SNe (which are unresolved
sources) if they are spherically symmetric:
any intrinsic polarization thus reveals asymmetry.
Core-collapse supernovae are, in fact, generally 
polarized in the continuum at levels of
$p\simeq 0.5$--$4$ \%\ due to electron scattering and their 
polarization increases after optical maximum light
(e.g., \citealt{jef91b,wan96,wan01,leo01b}); however, the polarization
falls to zero at very late times, when the electron scattering
opacity becomes very low (e.g., \citealt{jef91b}).
The typical line polarization profile---actually mostly
due to electron polarized light interacting with a line---predicted
theoretically \citep{jef89} and to some degree confirmed
observationally in SN~1987A and other supernovae
\citep{jef91a,jef91b,leo01a,leo01b},
is an inverted P~Cygni profile:
strong polarization maximum at the flux P~Cygni trough feature
and polarization minimum at the flux P~Cygni emission feature.
Line blending and other intrinsic effects can distort these
profiles.
For SN 1998bw, an intrinsic optical polarization of $0.4$--$0.6$ \%\
was found, suggesting an asymmetry of less than $2/1$ in the axial ratio
of the ejecta \citep{pat01,kay98}.
No distinct line polarization features were seen probably due to the
poor S/N or the relatively narrow wavelength range of
those observations.

\section{OBSERVATIONS AND DATA REDUCTION}

The spectropolarimetry was taken with the 8.2-m
Subaru Telescope equipped with the Faint Object Camera and
Spectrograph (FOCAS, \citealt{kas00,yos00})
from 2002 February 9 through June 9.
%The observing log is shown in Table \ref{tbl-1}.
For February and June observations, we used a 300 grooves mm$^{-1}$
grism (the central wavelength of 5500 \AA ) with a $0\farcs 4$ width
slit, resulting a spectral resolution ($\lambda/\Delta\lambda$) was
$\sim 1200$.
For March observation, we used another 300 gr mm$^{-1}$
grism (7500 \AA ) with a $0\farcs 8$ width slit, resulting in
$\lambda/\Delta\lambda \sim 650$.
The nightly total exposure time was 960 to 4000 sec.
The linear polarimetric module of FOCAS consists of a rotating
superachromatic half-wave plate and a crystal quartz Wollaston
prism, and both the ordinary and the extraordinary
rays are simultaneously recorded on two MIT/LL CCDs
(2k$\times$4k$\times$15\micron ).
A typical observing sequence consisted of four integrations
at the $\psi=0\arcdeg, 45\arcdeg, 22\fdg 5$ and $67\fdg 5$
positions of the half-wave plate.
Stokes $Q/I$ and $U/I$ were calculated as in \S 6.1.2 of \citet{tin96}.
For polarimetric calibration, we obtained data for unpolarized
and polarized standard stars, including measurements
of flatfield lamps through fully-polarizing filters.
%Although the stability of instrumental polarization and
%depolarization in FOCAS on the Subaru Telescope have not
%yet been fully calibrated, our results indicate that the
%instrumental polarization ($\lesssim 0.1$ \%) and
%the depolarization factor ($\lesssim 0.05$) are negligible at
%all wavelengths.
The flux was calibrated using observations of G191B2B and
BD+28$\arcdeg$4211 \citep{oke90},
and then was multiplied by a constant to match the
VSNET\footnote{\url{http://www/kusastro.kyoto-u.ac.jp/vsnet}}
photometric data.

\section{RESULTS AND DISCUSSION}

Figure \ref{fig1} shows the observed flux and polarization spectra.
Several blueshifted broad absorption lines can be identified
in the February flux spectrum \citep{maz02}.
For March and June flux spectra a detailed analysis
has yet to be done.
However, we note that the March spectra resemble that of
SN~1997ef at day 67 \citep{maz00}, including the onset of
net emission in \ion{Ca}{2} $\lambda\lambda$8498, 8542, 8662
(i.e., the \ion{Ca}{2} IR triplet).
The significant emission line at $\sim 6300$\AA\ in the June spectrum
is identified as [\ion{O}{1}] $\lambda$6300, 6363 as in
SN~1997ef \citep{maz01}.
The polarization is $\gtrsim 0.5$ \%\ at a position angle (PA) of
$\theta =120\arcdeg \pm 20\arcdeg$
over the observed wavelengths.
Significant day-by-day variation is not seen in the
polarization spectra within each month.
Here we will only analyze monthly averages.

%\subsection{Interstellar Polarization}
\subsection{Interstellar and Intrinsic Polarization}

Interstellar polarization (ISP) varies slowly with
wavelength in the optical and
is well approximated by the empirical formula
$p_{\rm ISP}(\lambda ) = p_{\rm max} \cdot \exp [ -1.15 \ln^{2}(
      \lambda_{\rm max}/\lambda) ]$,
where $p_{\rm max}=p_{\rm ISP}(\lambda_{\rm max})$ is the 
peak polarization (\citet{ser75}, hereafter SMF).
In March the emission feature of the \ion{Ca}{2} IR triplet line
profile shows strong net emission due to NLTE processes.
Such NLTE line flux is necessarily unpolarized on emission.
Since the line profile is still broad (absorption minimum at a
$\sim -14,000$ km s$^{-1}$ redshift, corresponding to an enclosed
mass of 2 M$_{\odot}$ in model CO100/4, which has a total
mass of 2.4 M$_{\odot}$ \citep{maz02}),
much of the emission is
probably coming from far out in the ejecta, where the
electron optical depth is low.
We conclude that the flux from this emission line is mostly
unscattered by electrons and unpolarized, and that it
dilutes the polarized electron scattered flux.
If this were the only effect, then the intrinsic polarization
should show a distinct minimum nearly exactly at the wavelength
of the flux emission maximum.
Since, in fact, the polarization is roughly constant
across the P~Cygni emission feature ($8400$--$9000$ \AA ),
apart from small variations that may be mostly noise,
the line is not only diluting the polarized
flux but is also probably strongly depolarizing it.
The intrinsic polarization across the emission feature
is probably close to zero.
We will assume that the observed polarization of the emission
feature is the ISP: thus
$p_{\rm ISP}(8600\;{\rm \AA})\approx 0.5$ \%.
We next adopt the median $\lambda_{\rm max}=5370$ \AA\ 
found by SMF for 30 stars with $p_{\rm max} / E_{B-V}\geq 7.0$.
Then from a non-linear regression we find
$p_{\rm max}=0.64\pm 0.20$ \%\ and
$\theta_{\rm ISP}=120\arcdeg\pm 10\arcdeg$.
(The uncertainties are crude estimates based on the
alternative assumption that only line flux dilution, and not
line depolarization, occurs in the region of the
\ion{Ca}{2} IR triplet emission feature.)
Since \citet{tak02} derive a color excess for SN~2002ap
of $E_{B-V}=0.09$ (a sum of $0.07$ within our Galaxy and
$0.02$ within M74) from interstellar Na D absorptions,
our assumption of the SMF $\lambda_{\rm max}$
is consistent: $p_{\rm max} / E_{B-V} = 0.64 / 0.09 \approx 7$.

The estimated ISP (EISP) is consistent with other factors.
In an ISP catalog \citep{hei00},
16 stars are within $10\arcdeg$ of SN~2002ap.
The data for these stars suggest a possible positive correlation
between polarization and the distance along the line
of sight toward SN~2002ap.
The two most distant stars among them, HD8919 ($d=525$ pc)
and HD9560 ($d=437$ pc) show
$(p,\ \theta) = (0.32\pm 0.10$ \mbox{ \%},
$99\arcdeg\pm 9\arcdeg)$ and
$(0.48\pm 0.09$ \mbox{ \%},
$123\arcdeg\pm 5\arcdeg)$, respectively.
On the other hand, it has been found that $p_{\rm max} (\% )$
has an empirical upper limit of $9E_{B-V}$ (SMF).
From the derived $E_{B-V} = 0.09$, an upper limit
on the ISP toward the supernova is $0.81$ \%.
The EISP is nicely sandwiched between the possible
lower bounds of the cited stars and the empirical upper limit.
The EISP position angle is also consistent with
the position angle of the spiral arm in M74 at the position
of SN~2002ap, $110\arcdeg$--$140\arcdeg$ (see DPOSS images).
The June 9 polarization spectrum has polarization consistent
with the EISP at the peak of the strong emission flux of the
[\ion{O}{1}] $\lambda$6300, 6363 forbidden line (Figure \ref{fig1}c):
one would expect that polarization to be ISP for a strong emission
line in the nebular epoch.
(The June 9 data has low S/N:  in the following, we analyze only the
higher S/N polarization spectra of the earlier two epochs.

%\subsection{Intrinsic Polarization}

Figure \ref{fig2} shows the intrinsic (i.e, EISP-subtracted) polarization
plotted on a {\it QU} diagram:
the polarization points are connected according
to their wavelength ordering.
%(The ISP correction was done by subtracting the ISP Stokes
%parameters from observed ones.)
Given the uncertainty in the EISP, points within
$0.2$ \%\ of the origin must be considered very uncertain.

If one assumes that the intrinsic supernova polarization is produced
by a single axisymmetric component in the ejecta, then
the intrinsic polarization plotted on a {\it QU} diagram
should lie on a line passing through the origin.
It can be seen that the polarization in February has two clear
position angles, PA less than or $\sim 120\arcdeg$ (associated
with the \ion{O}{1}/\ion{Ca}{2} line trough) and
PA$\sim 80\arcdeg \pm 20\arcdeg$ (associated with the continuum
from $\sim 5700$--$8200$ \AA), joined by a somewhat complicated
transition.
The polarization in March has a clear position angle
PA$\sim 150\arcdeg$ associated with the \ion{Ca}{2} line trough
and with at least some of the continuum.
We conclude that there are multiple asymmetric components,
and that their contribution varies with time.
It is likely that the recession of the supernova photosphere uncovers
different asymmetries.

\subsection{Possible Models}

Figures \ref{fig3}a, b, and c show the flux and 
intrinsic polarization corrected for
heliocentric redshift [$v_{\rm helio}=+631 \mbox{\rm km s}^{-1}$
\citep{sma02}] and interstellar extinction ($E_{B-V}=0.09$).
The figures show that polarization is low and barely significant
(given the uncertainty in the EISP), except in the regions
$\sim 6700$--$8000$ \AA\ for February and $\sim 6700$--$8300$
\AA\ for March.

The low continuum polarization blueward of
$\sim 6700$ \AA\ in both epochs may be due to the depolarizing
effect of lines \citep{how01}:
in supernova spectra, lines generally become stronger further
to the blue.
The continuum polarization, where it is significant
(a relatively small region) is $\sim 0.4$ \% in both epochs.
If the asymmetry is assumed to be an axisymmetric,
global prolate or oblate asymmetry, then 
the continuum polarization can be explained by
an axial ratio (assuming there is a main axis) projected on the sky
that is different from 1 by of order $10$ \%.
This estimate is a crude one based on realistic,
but parameterized, calculations \citep{hoe91,hoe95}.
The estimate is also crude because, as noted above, the
asymmetry cannot be completely axisymmetric.
The estimated asymmetry is not large, compared to those estimated for
some other supernovae (e.g., \citealt{wan01}).

The three distinct line polarization profiles seen in
Figure~\ref{fig3} (at the \ion{O}{1}/\ion{Ca}{2} flux absorption
in February and the \ion{O}{1} and \ion{Ca}{2} flux absorptions
in March) can partially be
accounted for by the inverted P~Cygni profile
(see \S~1): the polarization maxima associated with line
trough features are clear.
Without detailed modeling more information probably cannot be
extracted from these profiles.

For the February continuum polarization, we suggest a
radically different origin from the bulk asymmetry
assumed in prolate/oblate
models or element inhomogeneity models (see below).
In Figure~\ref{fig3}d we show the intrinsic polarized flux ($p\times F$)
compared to the observed flux scaled down by a factor of $0.0018$ and
non-relativistically redshifted by a velocity of $0.23c$:
i.e., $\lambda_{\rm redshifted}=\lambda/0.77$.
There is fair agreement over the range $\sim 5000$--$8000$ \AA.
In response to this comparison, \citet{leo02} made
a similar comparison for their SN 2002ap spectropolarimetry and
also obtained a similarly good agreement.
The agreement suggests (but does not prove)
that a large component of polarized flux comes from
electron scattering in an ionized clump (i.e., a jet)
thrown out of the supernova explosion.
(We assume a single jet or clump for simplicity here, although
a pair of bipolar jets are a physical possibility 
(e.g., \citealt{whe02}).)
In a simple non-relativistic picture,
the scattered light is redshifted by
$v_{\rm red}\sim v_{\rm jet}(1+\cos i)$, where,
$v_{\rm red}=0.23c$, $v_{\rm jet}$
is the characteristic velocity of scattering relative to
the supernova center, and $i$ is the inclination
angle of jet to the line of sight measured from the far side
of the supernova.
The jet polarization component is calculated from
\[
p_{\rm jet}(\lambda) =
f\cdot\frac{F[\lambda(1-v_{\rm red}/c)]}{F(\lambda)} \mbox{ ,}
\]
where $F(\lambda)$ is the corrected flux,
$1-v_{\rm red}/c$ is the blueshift back to the origin of
the scattered flux observed at $\lambda$, and the scale
factor $f=0.0018$:  $f$ accounts both for
the polarization of the jet scattered flux and the fraction scattered.
The polarization is wavelength dependent even though
electron scattering is wavelength-independent since the scattered
flux comes from a bluer part of the spectrum than the
part it is added to.
Electron scattering depends on scattering direction:
e.g., maximum polarization occurs for $i=90^{\circ}$;
half as much for $i=45^{\circ}$ or $135^{\circ}$;
zero for $i=0^{\circ}$ or $180^{\circ}$.
The jet velocities corresponding to $i=90^{\circ}$,
$45^{\circ}$, and $0^{\circ}$ are
$0.23c$, $0.135c$, and $0.115c$, respectively.
Thus $0.115c$ is a lower bound on the jet velocity
and $i\gtrsim 90^{\circ}$ would require a somewhat
relativistic jet.

High velocity jet-like clumps have been
proposed in some hydrodynamic explosion models for SNe and GRB's
(e.g., \citealt{nag97,mac99,mae02,whe02}).
If a jet is thrown out of an exploding supernova core,
then it is plausible that it carries some radioactive $^{56}$Ni.
The gamma-rays from decay would keep the jet ionized to
some degree just as they keep the nebular phase bulk ejecta ionized.

If the jet picture is correct, then the position angle
of the jet on the sky is $\sim 170^{\circ}$ (or $\sim 350^{\circ}$)
since the jet polarization component has position angle $\sim 80^{\circ}$
and electron scattering polarizes perpendicularly to the
scattering plane.
The \ion{O}{1}/\ion{Ca}{2} line polarization maximum in the February
data cannot easily be associated with the jet.
As Figure~2 shows, the observed position angle makes
an excursion from $\sim 80^{\circ}$
up to $120^{\circ}$ across the polarization maximum.
Some of the line polarization may arise
in the bulk asymmetry of the supernova.
It is possible that the position angle of
$\sim 120^{\circ}$ is the net result of a jet polarizing at
$\sim 80^{\circ}$ and a bulk asymmetry polarizing at
$\sim 150^{\circ}$ (i.e., at the position angle
observed in the March data).
To test this model we have eliminated the jet polarization component
from the February intrinsic polarization by subtracting
jet polarization Stokes parameters from the intrinsic polarization
Stokes parameters.
The residual polarization and position angle spectra are
plotted in Figures \ref{fig3}e,f.
The position angle of the residual polarization for
February from the region of significant polarization
(i.e., $\sim 6700$--$8000$ \AA) is now approximately
centered on $150^{\circ}$ and deviates by more than
$30^{\circ}$ only in a few isolated points.
The results in Figure~\ref{fig3}f are thus consistent with
the jet model.

The jet may not be a completely separated amount of ejecta,
but rather a blob rich in $^{56}$Ni moving at $\gtrsim 0.115c$.
A high-velocity, $^{56}$Ni-rich region is required both in the
hypernova explosion models of SN 1998bw \citep{nak01,mae02},
and in SN~2002ap
\citep{maz02} in order to reproduce the light curves.
Spectral synthesis for SN~2002ap \citep{maz02} suggests
some material at velocities higher than the photospheric 
velocities of $\sim 0.1c$--$0.117c$ \citep{maz02,kin02} up to $0.22c$.
The ionization of the blob would likely be increased by radioactive
$^{56}$Ni and this would likely make the blob more polarizing
than other parts of the ejecta at the same velocity leading to
net polarization.
Other chemical inhomogeneities in the ejecta at varying velocities
are also possible \citep{mae02} and would affect polarization in
complicated ways.

\section{CONCLUDING REMARKS}

Our results are summarized in the Abstract.
%In this Letter we have presented spectropolarimetry for
%SN~2002ap and given its first order interpretation.
%We suggest that the supernova has a bulk asymmetry with
%an axial ratio projected on the sky that is different from 1
%by of order $10$ \% and speculatively a single polarizing jet
%moving at $\gtrsim 0.115c$.
%The jet can make a significant contribution to the
%polarization only in the February observational epoch.
More realistic modeling is necessary for a more
definitive understanding of the polarization.
The degree of the bulk asymmetry suggested in this paper may be tested
with the line widths and their ratios in the nebular spectrum
\citep{mae02}.

\acknowledgments

We are grateful to the staff members at the Subaru Telescope
for their kind help and their rearrangement of the telescope
maintenance schedule for our observation on March 8.
This work has been supported in part by the grant-in-Aid
for Scientific Research (12640233, 14047206, 14540223) and
COE research (07CE2002) of the Ministry of Education, Science,
Culture, Sports, and Technology in Japan.

\clearpage

%{\scriptsize
%\begin{deluxetable}{lccrr}
%\tabletypesize{\scriptsize}
%%\rotate
%\tablenum{1}
%\tablewidth{0pt}
%\tablecaption{Log of Observations for SN~2002ap\label{tbl-1}}
%\tablehead{
%\colhead{Date (UT)} &
%\colhead{Grism\tablenotemark{a}} &
%\colhead{$\lambda\lambda$ (\AA )\tablenotemark{b}} &
%\colhead{$\lambda/\Delta\lambda$} &
%\colhead{$\Delta t$ (s)}
%}
%\startdata
%2002 Feb  9.2 & 300/5500 & 4750--8300 & 1200 & $1200$ \\
%2002 Feb  9.3 & 300/5500 & 3850--6050 & 1200 & $2800$ \\
%2002 Feb 10.3 & 300/5500 & 4750--8300 & 1200 & $1200$ \\
%2002 Feb 11.3 & 300/5500 & 4750--8300 & 1200 & $1600$ \\
%2002 Feb 11.3 & 300/5500 & 3850--6050 & 1200 & $1600$ \\
%2002 Feb 12.3\tablenotemark{c} & 300/5500 & 4750--8300 & 1200 & $1200$ \\
%2002 Mar  8.2 & 300/7500 & 4850--9050 &  650 &  $960$ \\
%2002 Mar 10.2 & 300/7500 & 4850--9050 &  650 & $1080$ \\
%2002 Jun  9.6 & 300/5500 & 4750--8300 & 1200 & $1200$ \\
%\tablenotetext{a}{Grooves per millimeter/central wavelength in angstroms}
%\tablenotetext{b}{Effective wavelength range of the observation, which
%depends on the combination of the grism and the order-cut filter used.}
%\tablenotetext{c}{On February 12 we could not carry out the whole sequence of
%polarimetry because of unstable weather, and so obtained only flux data.}
%\tablerefs{}
%\tablecomments{}
%\enddata
%\end{deluxetable}
%}
%
%\clearpage

\figcaption[Kawabata.fig1.eps]{Flux and polarization spectra of SN
2002ap. Heliocentric redshift, interstellar extinction and
polarization have not been corrected for.
 From top to bottom, we plot (a) total flux, (b, c) polarization
level $p$ and position angle $\theta$ on each observation day
as indicated.
The polarimetric data are binned to a constant photon noise
of $0.05$ \%
which is shown by the error bars of polarization points.
The EISP component is shown by a dashed curve in (b,c).
\label{fig1}}

\figcaption[Kawabata.fig2.eps]{{\it QU}-diagram of the monthly-averaged
intrinsic (i.e., ISP-subtracted) 
polarization for February and March epochs.
The data are binned to
a constant photon noise of $0.04$ \%.
It can be seen that the polarization in February has, at least, two
preferred axes: PA$\sim 120\arcdeg$ (associated with the
\ion{O}{1}/\ion{Ca}{2} line trough) and PA$\sim 80\arcdeg$
(associated with the continuum).
The polarization in March has a clear position angle PA$\sim 150\arcdeg$
associated with the \ion{Ca}{2} line trough and the significantly
polarized continuum.
These position angles are indicated by thick arrows.
Note that the position angle on the sky is half the angular
location on a {\it QU} diagram.
\label{fig2}}

\figcaption[Kawabata.fig3.eps]{Intrinsic polarization
spectra corrected for heliocentric redshift and interstellar extinction.
From top to bottom, we plot (a) total flux in erg s$^{-1}$ cm$^{-2}$
\AA$^{-1}$, (b) polarization level $p$, (c) position angle $\theta$,
(d) polarized flux, and (e,f) $p$ and $\theta$ of the residual
polarization after the jet polarization component has been
subtracted from the February data.
The February flux is the mean of Feb 9 and 11, and the March
flux is the mean of Mar 8 and 10.
We adopt
% a heliocentric redshift $+631$ km s$^{-1}$ for M74
%(Smartt \& Meikle 2002), a color excess of $E_{B-V}=0.07$
%in our Galaxy and $0.02$ in M74 (Takada-Hidai et al. 2002) and
the normal interstellar extinction curve (Cardelli, Clayton,
\& Mathis 1989).
Deep absorption bands due to the terrestrial atmosphere
and the interstellar medium have been removed by interpolation
using nearby continuum levels.
The polarimetric data are binned in the same manner as in Fig. 2.
The solid curve in (d) is the February flux multiplied by
$0.0018$ and redshifted by $+0.23c$ (see \S 3.2).
\label{fig3}}

\plotone{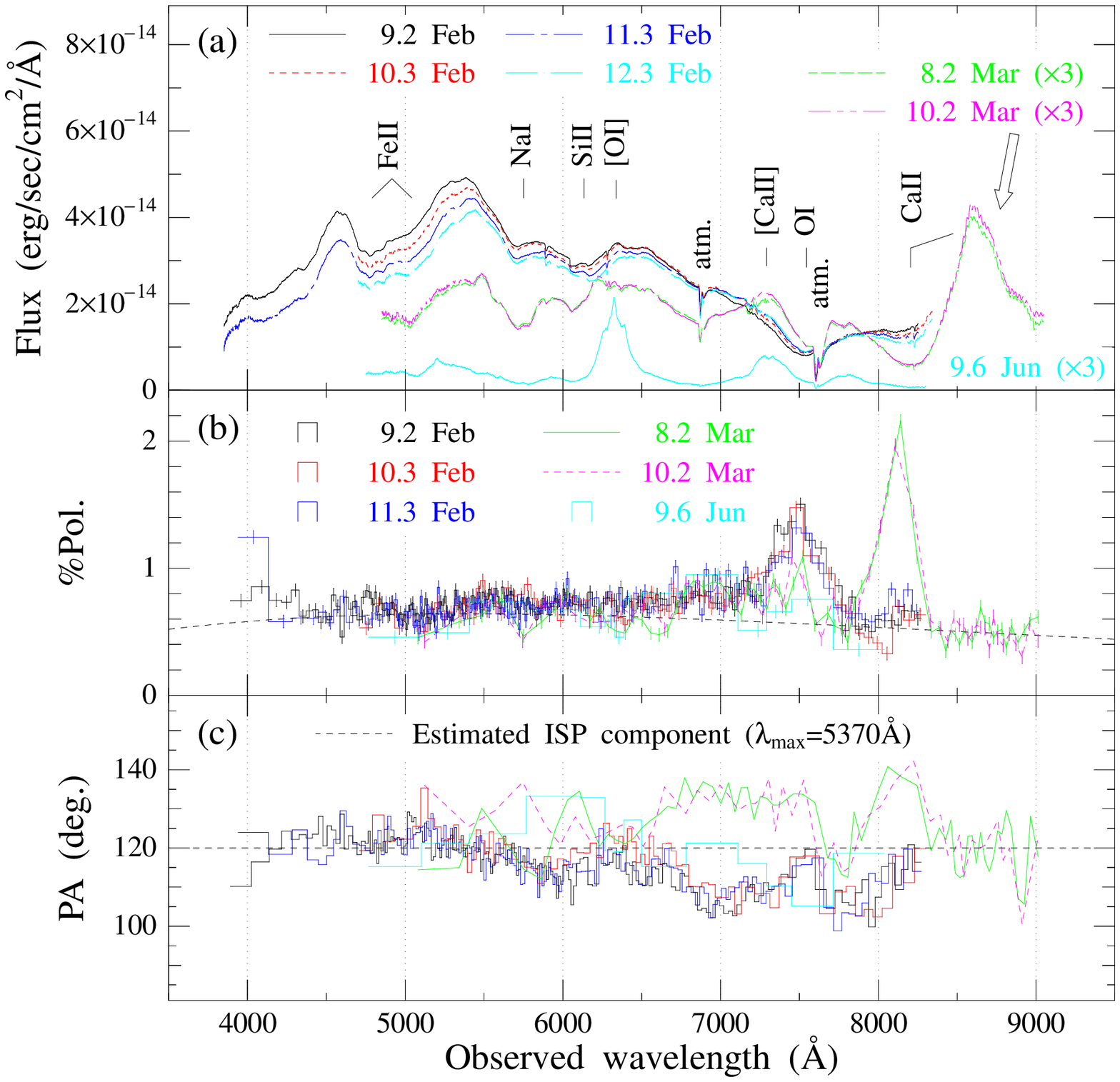}
\plotone{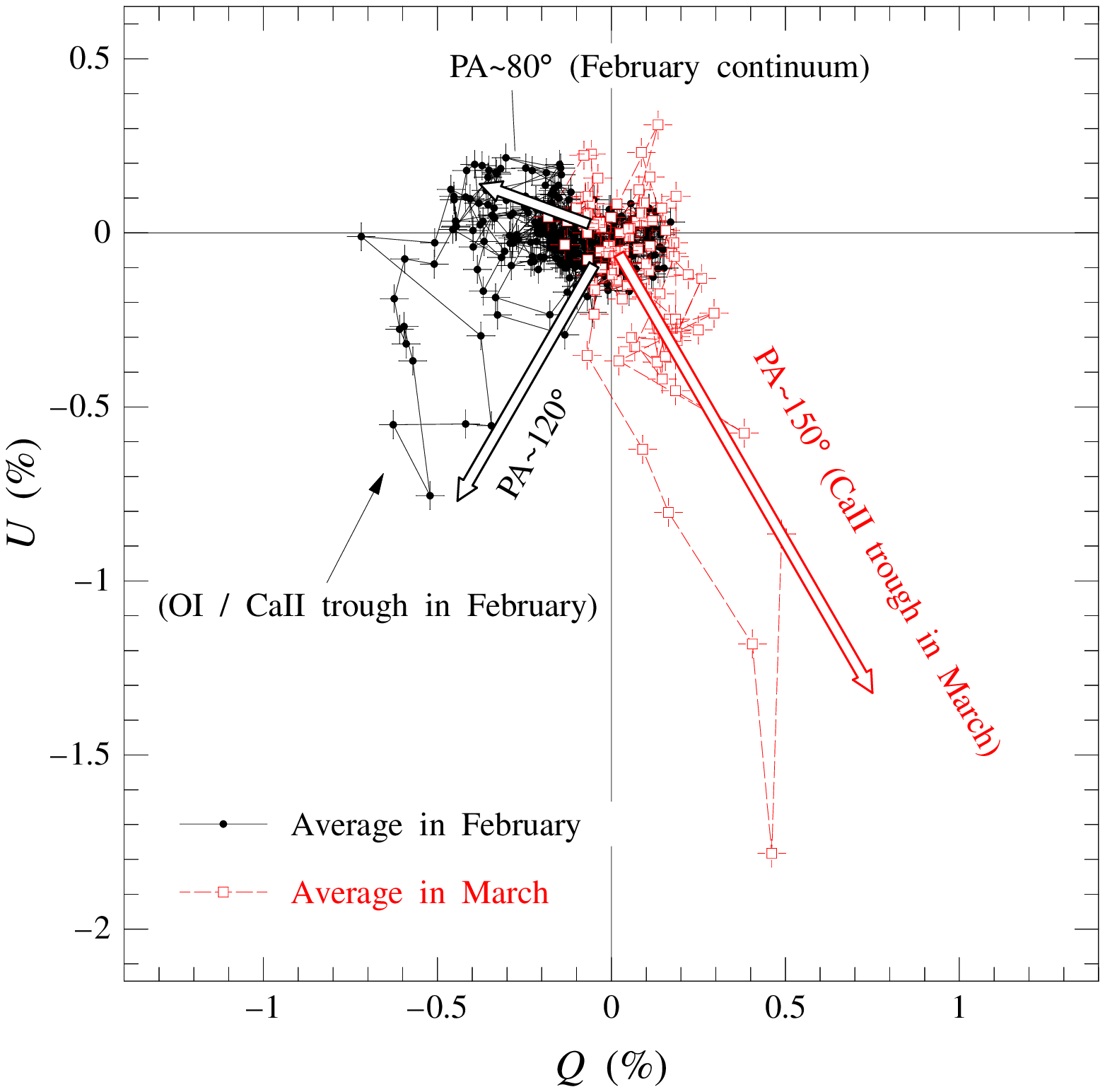}
\plotone{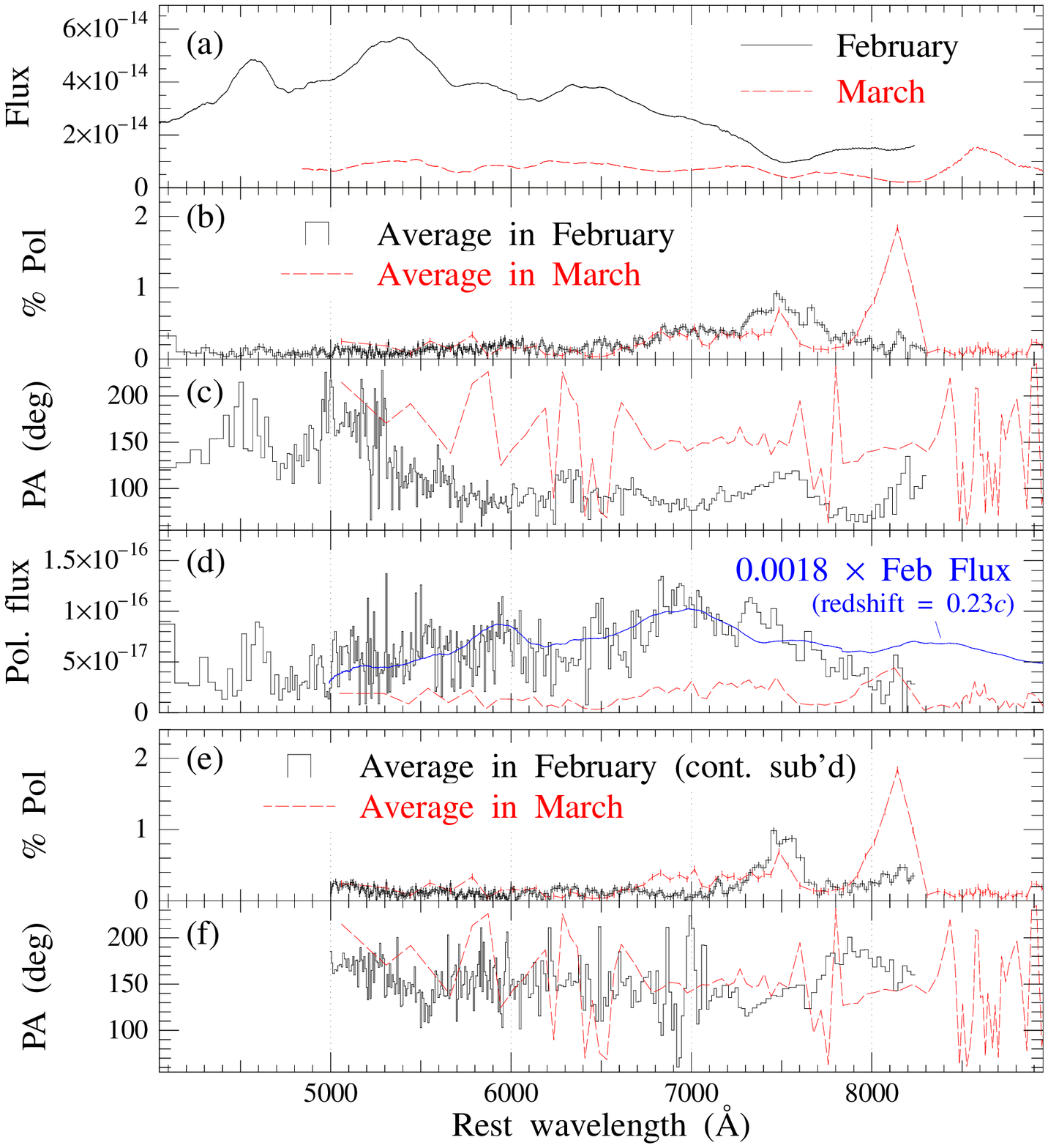}

%\clearpage

\end{document}